\documentclass[preprint,review,12pt]{elsarticle} 
\biboptions{sort&compress}
\usepackage{booktabs}
\usepackage{amsmath}
\begin{document}
\begin{frontmatter}

\title{Renormalization group approach to vibrational energy transfer in protein}

\author{Shigenori Tanaka}
\address{Graduate School of System Informatics, Department of Computational Science, \\
Kobe University, 1-1 Rokkodai, Nada-ku, Kobe 657-8501, Japan}
\ead{tanaka2@kobe-u.ac.jp; Fax:+81-78-803-6621}

\begin{abstract}

Renormalization group method is applied to the study of vibrational energy transfer in protein molecule.
An effective Lagrangian and associated equations of motion to describe the resonant energy transfer are analyzed in terms of the first-order 
perturbative renormalization group theory that has been developed as a unified tool for global asymptotic analysis.
After the elimination of singular terms associated with the Fermi resonance, amplitude equations to describe the slow dynamics of vibrational 
energy transfer are derived, which recover the result obtained by a technique developed in nonlinear optics 
[S.J. Lade, Y.S. Kivshar, Phys. Lett. A 372 (2008) 1077].

\end{abstract}

\end{frontmatter}

\newpage

\section{Introduction} 

The study of vibrational motions in protein system has attracted much interest in the context of their relations to biological functions, 
paying attention to the intrinsic flexibility of large biomolecules.
A lot of theoretical investigations on the basis of the normal mode or principal component analysis 
\cite{Cui,Ama,Hay,Kitao,Tanaka2} have been carried out 
in combination with molecular mechanics or dynamics simulations for protein systems, thus providing useful information at molecular level.
Vibrational energy transfers between the normal modes are then expected when the nonlinear, anharmonic couplings are taken into accout \cite{Piazza,Gur}.
The dynamical relaxation behavior after the excitation of vibrational modes is thus central for a deep understanding of 
the relationships among structure, dynamics and function in protein systems.

\par

In this context, Moritsugu et al. \cite{Moritsugu1,Moritsugu2} studied the intramolecular transfer of vibrational energy in myoglobin 
by means of molecular dynamics simulation.
They found that the vibrational energy was transferred from a given normal mode to very few other modes that were selected by the 
frequency relation of the Fermi resonance \cite{Uzer} and the magnitude of the third-order mode coupling term.
It was also observed that the magnitude of the coupling coefficients was mainly determined by the degree of geometrical overlap of the 
corresponding normal modes.
Further, performing a molecular dynamics simulation with just these few modes, they reproduced the results for all-atom simulation well, thus 
confirming that only these modes were important for the dynamics of vibrational energy transfer.

\par

As is well known, the description by naive perturbation theory breaks down for this kind of resonant energy transfer problem due to the 
emergence of singular terms associated with the Fermi resonance \cite{Uzer}.
Invoked by a theoretical technique for describing the resonant parametric interaction between different harmonics in nonlinear optics, 
Lade and Kivshar \cite{Lade} proposed a simple theory to overcome this difficulty, which provided coupled mode equations for 
slowly varying envelopes of the vibrational modes.
They thus found that their equations could describe the dynamics of the resonant energy transfer between the vibrational modes observed 
in the simulations by Moritsugu et al. \cite{Moritsugu1,Moritsugu2} with good accuracy, and gave a wealth of suggestions about the underlying mechanism of 
nonlinear dynamics of vibrational energy transfer \cite{Tanaka}.

\par

In this work, the renormalization group (RG) approach which has been developed as a unified tool for global asymptotic analysis \cite{Chen,Nozaki,Kirk} is 
employed in order to describe the long-time dynamics of the resonant vibrational energy transfer in protein system.
Starting with the same model Lagrangian as studied by Moritsugu et al. \cite{Moritsugu1,Moritsugu2} and Lade et al. \cite{Lade}, we apply the RG 
technique at the first-order perturbation level and derive the corresponding amplitude equations through the elimination of singular terms associated 
with the Fermi resonance.
We then find that the derived equations are essentially of the same form as those obtained by Lade and Kivshar \cite{Lade} at the lowest-order level.
In the following section, the theoretical formulation of the present RG approach is addressed. 
Some discussions and conclusion are given in Sec. 3.

\section{Theory}

We start with a model Lagrangian given by Moritsugu et al. \cite{Moritsugu1,Moritsugu2}, 

\begin{equation}
L = \sum_{j=1}^{4}\frac{1}{2}\dot{q}_{j}^{2} - \sum_{j=1}^{4}\frac{1}{2}\omega_{j}^{2}q_{j}^{2} -\alpha q_{1}q_{2}q_{3} -\beta q_{1}^{2}q_{4},
\end{equation}

\noindent
where $q_{j}$ refers to the coordinate of the $j$th normal mode with the frequency $\omega_{j}$ and the dot denotes the time derivative.
Two types of third-order coupling among the modes are considered in the model, which are characterized by small coupling parameters $\alpha$ and $\beta$.
Equations of motion for $q_{j}$ then read 

\begin{equation}
\ddot{q}_{1} = -\omega_{1}^{2}q_{1} -\alpha q_{2}q_{3} -2\beta q_{1}q_{4}, \nonumber
\end{equation}

\begin{equation}
\ddot{q}_{2} = -\omega_{2}^{2}q_{2} -\alpha q_{1}q_{3}, \nonumber
\end{equation}

\begin{equation}
\ddot{q}_{3} = -\omega_{3}^{2}q_{3} -\alpha q_{1}q_{2}, \nonumber
\end{equation}

\begin{equation}
\ddot{q}_{4} = -\omega_{4}^{2}q_{4} -\beta q_{1}^{2}. 
\end{equation}

\noindent
These equations of motion were found to appropriately reproduce the temporal variations of vibrational energy transfer observed in the all-atom 
molecular dynamics simulation for myoglobin \cite{Moritsugu1,Moritsugu2}.

\par

The solution to $q_{j}$ is given by a perturbation theory up to the first order of $\alpha$ and $\beta$ as 

\begin{eqnarray}
q_{1}(t) &=& A_{1}e^{i\omega_{1}t} - \frac{\alpha A_{2}A_{3}}{\omega_{1}^{2}-(\omega_{2}+\omega_{3})^{2}}e^{i(\omega_{2}+\omega_{3})t} 
- \frac{\alpha A_{2}^{*}A_{3}}{\omega_{1}^{2}-(\omega_{3}-\omega_{2})^{2}}e^{i(\omega_{3}-\omega_{2})t} \nonumber \\
&-& \frac{2\beta A_{1}A_{4}}{\omega_{1}^{2}-(\omega_{1}+\omega_{4})^{2}}e^{i(\omega_{1}+\omega_{4})t}
- \frac{2\beta A_{1}^{*}A_{4}}{\omega_{1}^{2}-(\omega_{4}-\omega_{1})^{2}}e^{i(\omega_{4}-\omega_{1})t} + \rm{c.c.}, \nonumber
\end{eqnarray}

\begin{equation}
q_{2}(t) = A_{2}e^{i\omega_{2}t} - \frac{\alpha A_{1}A_{3}}{\omega_{2}^{2}-(\omega_{1}+\omega_{3})^{2}}e^{i(\omega_{1}+\omega_{3})t} 
- \frac{\alpha A_{1}A_{3}^{*}}{\omega_{2}^{2}-(\omega_{1}-\omega_{3})^{2}}e^{i(\omega_{1}-\omega_{3})t} + \rm{c.c.}, \nonumber
\end{equation}

\begin{equation}
q_{3}(t) = A_{3}e^{i\omega_{3}t} - \frac{\alpha A_{1}A_{2}}{\omega_{3}^{2}-(\omega_{1}+\omega_{2})^{2}}e^{i(\omega_{1}+\omega_{2})t} 
- \frac{\alpha A_{1}A_{2}^{*}}{\omega_{3}^{2}-(\omega_{1}-\omega_{2})^{2}}e^{i(\omega_{1}-\omega_{2})t} + \rm{c.c.}, \nonumber
\end{equation}

\begin{equation}
q_{4}(t) = A_{4}e^{i\omega_{4}t} - \frac{\beta A_{1}^{2}}{\omega_{4}^{2}-4\omega_{1}^{2}}e^{2i\omega_{1}t} 
- \frac{\beta A_{1}A_{1}^{*}}{\omega_{4}^{2}} + \rm{c.c.},
\end{equation}

\noindent
where we have introduced amplitude parameters, $A_{j}$, and c.c.\ (or the asterisk) means the complex conjugate.

\par

Here, we consider a case of vibrational energy transfer in which the Fermi resonance takes place.
According to Moritsugu et al. \cite{Moritsugu1,Moritsugu2}, it is assumed that two frequency parameters, 
$\Omega_{\alpha}=\omega_{1}-\omega_{2}-\omega_{3}$ and $\Omega_{\beta}=2\omega_{1}-\omega_{4}$, satisfy such a condition as 
$\vert\Omega_{\alpha}\vert, \vert\Omega_{\beta}\vert \ll \omega_{j}$.
Retaining only the terms of the first order of $\alpha$ or $\beta$ that become large under the Fermi resonance condition on the 
right-hand side of Eq.\ (3), we find 

\begin{equation}
q_{1}(t) \approx \left[A_{1} - \frac{\alpha A_{2}A_{3}}{\Omega_{\alpha}(\omega_{1}+\omega_{2}+\omega_{3})}e^{-i\Omega_{\alpha}t} 
- \frac{2\beta A_{1}^{*}A_{4}}{\Omega_{\beta}\omega_{4}}e^{-i\Omega_{\beta}t}\right]e^{i\omega_{1}t} + \rm{c.c.}, \nonumber
\end{equation}

\begin{equation}
q_{2}(t) \approx \left[A_{2} + \frac{\alpha A_{1}A_{3}^{*}}{\Omega_{\alpha}(\omega_{1}+\omega_{2}-\omega_{3})}e^{i\Omega_{\alpha}t}\right]e^{i\omega_{2}t} + \rm{c.c.}, \nonumber
\end{equation}

\begin{equation}
q_{3}(t) \approx \left[A_{3} + \frac{\alpha A_{1}A_{2}^{*}}{\Omega_{\alpha}(\omega_{1}-\omega_{2}+\omega_{3})}e^{i\Omega_{\alpha}t}\right]e^{i\omega_{3}t} + \rm{c.c.}, \nonumber
\end{equation}

\begin{equation}
q_{4}(t) \approx \left[A_{4} + \frac{\beta A_{1}^{2}}{\Omega_{\beta}(2\omega_{1}+\omega_{4})}e^{i\Omega_{\beta}t}\right]e^{i\omega_{4}t} + \rm{c.c.}. 
\end{equation}

\par

According to the RG technique \cite{Chen,Nozaki,Kirk}, 
we express the amplitude parameters $A_{j}$ in Eq.\ (4) up to the first order of $\alpha$ or $\beta$ as 

\begin{equation}
A_{1} = \tilde{A}_{1}(\tau)\left[1+\alpha Z_{1\alpha}(\tau)+\beta Z_{1\beta}(\tau)\right], \nonumber
\end{equation}

\begin{equation}
A_{2} = \tilde{A}_{2}(\tau)\left[1+\alpha Z_{2\alpha}(\tau)\right], \nonumber
\end{equation}

\begin{equation}
A_{3} = \tilde{A}_{3}(\tau)\left[1+\alpha Z_{3\alpha}(\tau)\right], \nonumber
\end{equation}

\begin{equation}
A_{4} = \tilde{A}_{4}(\tau)\left[1+\beta Z_{4\beta}(\tau)\right], 
\end{equation}

\noindent
where we have introduced a time variable $\tau$ and renormalization constants $Z_{j\alpha}$, $Z_{j\beta}$.
In order to eliminate the singular terms in Eq.\ (4) up to the lowest order, the renormalization constants are required to satisfy the following equations: 

\begin{equation}
\tilde{A}_{1}(\tau)Z_{1\alpha}(\tau) = \frac{\tilde{A}_{2}(\tau)\tilde{A}_{3}(\tau)e^{-i\Omega_{\alpha}\tau}}{\Omega_{\alpha}(\omega_{1}+\omega_{2}+\omega_{3})}, \nonumber
\end{equation}

\begin{equation}
\tilde{A}_{1}(\tau)Z_{1\beta}(\tau) = \frac{2\tilde{A}_{1}^{*}(\tau)\tilde{A}_{4}(\tau)e^{-i\Omega_{\beta}\tau}}{\Omega_{\beta}\omega_{4}}, \nonumber
\end{equation}

\begin{equation}
\tilde{A}_{2}(\tau)Z_{2\alpha}(\tau) = -\frac{\tilde{A}_{1}(\tau)\tilde{A}_{3}^{*}(\tau)e^{i\Omega_{\alpha}\tau}}{\Omega_{\alpha}(\omega_{1}+\omega_{2}-\omega_{3})}, \nonumber
\end{equation}

\begin{equation}
\tilde{A}_{3}(\tau)Z_{3\alpha}(\tau) = -\frac{\tilde{A}_{1}(\tau)\tilde{A}_{2}^{*}(\tau)e^{i\Omega_{\alpha}\tau}}{\Omega_{\alpha}(\omega_{1}-\omega_{2}+\omega_{3})}, \nonumber
\end{equation}

\begin{equation}
\tilde{A}_{4}(\tau)Z_{4\beta}(\tau) = -\frac{\tilde{A}_{1}(\tau)^{2}e^{i\Omega_{\beta}\tau}}{\Omega_{\beta}(2\omega_{1}+\omega_{4})}.
\end{equation}

\par

Since the amplitude parameters $A_{j}$ should not depend on the introduced time variable $\tau$, we call for identities as

\begin{equation}
\frac{dA_{j}}{d\tau} = 0.
\end{equation}

\noindent
We then obtain the RG equations \cite{Chen,Nozaki,Kirk} up to the first order of $\alpha$ or $\beta$ as 

\begin{equation}
\frac{d\tilde{A}_{1}(\tau)}{d\tau} + \left[\alpha\frac{dZ_{1\alpha}(\tau)}{d\tau}+\beta\frac{dZ_{1\beta}(\tau)}{d\tau}\right]\tilde{A}_{1}(\tau) = 0, \nonumber
\end{equation}

\begin{equation}
\frac{d\tilde{A}_{2}(\tau)}{d\tau} + \alpha\frac{dZ_{2\alpha}(\tau)}{d\tau}\tilde{A}_{2}(\tau) = 0, \nonumber
\end{equation}

\begin{equation}
\frac{d\tilde{A}_{3}(\tau)}{d\tau} + \alpha\frac{dZ_{3\alpha}(\tau)}{d\tau}\tilde{A}_{3}(\tau) = 0, \nonumber
\end{equation}

\begin{equation}
\frac{d\tilde{A}_{4}(\tau)}{d\tau} + \beta\frac{dZ_{4\beta}(\tau)}{d\tau}\tilde{A}_{4}(\tau) = 0,
\end{equation}

\noindent
from Eq.\ (5), considering 
$d\tilde{A}_{j}(\tau)/d\tau \sim O(\alpha)$ or $O(\beta)$.
We also note the lowest-order relations as

\begin{equation}
\tilde{A}_{1}(\tau)\frac{dZ_{1\alpha}(\tau)}{d\tau} = -\frac{i\tilde{A}_{2}(\tau)\tilde{A}_{3}(\tau)e^{-i\Omega_{\alpha}\tau}}{\omega_{1}+\omega_{2}+\omega_{3}}, \nonumber
\end{equation}

\begin{equation}
\tilde{A}_{1}(\tau)\frac{dZ_{1\beta}(\tau)}{d\tau} = -\frac{2i\tilde{A}_{1}^{*}(\tau)\tilde{A}_{4}(\tau)e^{-i\Omega_{\beta}\tau}}{\omega_{4}}, \nonumber
\end{equation}

\begin{equation}
\tilde{A}_{2}(\tau)\frac{dZ_{2\alpha}(\tau)}{d\tau} = -\frac{i\tilde{A}_{1}(\tau)\tilde{A}_{3}^{*}(\tau)e^{i\Omega_{\alpha}\tau}}{\omega_{1}+\omega_{2}-\omega_{3}}, \nonumber
\end{equation}

\begin{equation}
\tilde{A}_{3}(\tau)\frac{dZ_{3\alpha}(\tau)}{d\tau} = -\frac{i\tilde{A}_{1}(\tau)\tilde{A}_{2}^{*}(\tau)e^{i\Omega_{\alpha}\tau}}{\omega_{1}-\omega_{2}+\omega_{3}}, \nonumber
\end{equation}

\begin{equation}
\tilde{A}_{4}(\tau)\frac{dZ_{4\beta}(\tau)}{d\tau} = -\frac{i\tilde{A}_{1}(\tau)^{2}e^{i\Omega_{\beta}\tau}}{2\omega_{1}+\omega_{4}},
\end{equation}

\noindent
from Eq.\ (6).
Combining Eqs.\ (8) and (9), we finally obtain differential equations, 

\begin{equation}
\frac{d\tilde{A}_{1}(\tau)}{d\tau} = \frac{i\alpha\tilde{A}_{2}(\tau)\tilde{A}_{3}(\tau)e^{-i\Omega_{\alpha}\tau}}{\omega_{1}+\omega_{2}+\omega_{3}} 
+ \frac{2i\beta\tilde{A}_{1}^{*}(\tau)\tilde{A}_{4}(\tau)e^{-i\Omega_{\beta}\tau}}{\omega_{4}}, \nonumber
\end{equation}

\begin{equation}
\frac{d\tilde{A}_{2}(\tau)}{d\tau} = \frac{i\alpha\tilde{A}_{1}(\tau)\tilde{A}_{3}^{*}(\tau)e^{i\Omega_{\alpha}\tau}}{\omega_{1}+\omega_{2}-\omega_{3}}, \nonumber
\end{equation}

\begin{equation}
\frac{d\tilde{A}_{3}(\tau)}{d\tau} = \frac{i\alpha\tilde{A}_{1}(\tau)\tilde{A}_{2}^{*}(\tau)e^{i\Omega_{\alpha}\tau}}{\omega_{1}-\omega_{2}+\omega_{3}}, \nonumber
\end{equation}

\begin{equation}
\frac{d\tilde{A}_{4}(\tau)}{d\tau} = \frac{i\beta\tilde{A}_{1}(\tau)^{2}e^{i\Omega_{\beta}\tau}}{2\omega_{1}+\omega_{4}},
\end{equation}

\noindent
to determine the renormalized amplitudes $\tilde{A}_{j}(\tau)$.

\par

We now equate the time variable $\tau$ with time $t$ in Eq.\ (10), and solving the differential equations with the initial condition of 
$q_{j}(0)=\tilde{A}_{j}(0)+\tilde{A}_{j}^{*}(0)$, we can obtain $\tilde{A}_{j}(t)$.
Thus, we find a solution to $q_{j}$ as 

\begin{equation}
q_{j}(t) = \tilde{A}_{j}(t)e^{i\omega_{j}t} + \tilde{A}_{j}^{*}(t)e^{-i\omega_{j}t}.
\end{equation}

\noindent
The energy of each mode can then be evaluated in the harmonic approximation \cite{Moritsugu1,Moritsugu2,Lade}.

\section{Discussion and Conclusion} 

In the preceding section we have derived an amplitude equation, Eq.\ (10), on the basis of the RG approach.
This equation is free from the singularities associated with the Fermi resonance, which have been observed in the naive perturbation theory, and 
describes a nonlinear, slow dynamics of vibrational energy transfer \cite{Tanaka} in terms of the renormalized amplitudes $\tilde{A}_{j}$.
The present renormalization approach has thus succeeded in an effective separation of the two time domains associated with the fast dynamics governed by $\omega_{j}$ 
and the slow (``envelope") dynamics for vibrational energy transfer in protein.

\par

Lade and Kivshar \cite{Lade} analyzed the same model as in the present study, and derived an analogous set of amplitude equations by integrating out the fast vibrational motions 
in an intuitive way developed in the field of nonlinear optics.
By solving the resultant equations numerically, they found that the long-time behavior of the vibrational energy transfer in the model Eq.\ (1) can be described 
very well, thus demonstrating the excellence of their method.
Taking into account the condition that $\vert\Omega_{\alpha}\vert$ and $\vert\Omega_{\beta}\vert \ll \omega_{j}$, their derived equations are 
essentially identical to the present Eq.\ (10).
That is, their result can be recovered in terms of the first-order perturbation theory in the RG approach, and can be systematically improved 
in the more general framework of the present theory.

\par

In order to explain the dynamical behavior of vibrational energy transfer observed in their all-atom molecular dynamics simulation for myoglobin, 
Moritsugu et al. \cite{Moritsugu1,Moritsugu2} considered a model system given by Eq.\ (1).
Performing a molecular dynamics simulation for the model system with the parameter set of $\omega_{1}=5.385$, $\omega_{2}=2.362$, $\omega_{3}=3.024$, 
$\omega_{4}=10.779$, $\alpha=-0.13$ and $\beta=0.12$, they reproduced the temporal evolution of the resonant energy transfers among the normal modes quantitatively.
Lade and Kivshar \cite{Lade} then solved their amplitude equations numerically with the same parameter set as that by Moritsugu et al., and found 
a good agreement between the two results for the time course of energy variations of the normal modes.
The energy transfer dynamics observed in their studies is essentially a renormalized one in the sense that a slow dynamics emerges in contrast to the normal-mode 
dynamics characterized by $\omega_{j}$.
The present work reproduces the result by Lade and Kivshar at the lowest-order level, and provides a more general framework for describing the resonant vibrational 
energy transfer in protein on the basis of the RG theory.
This line of theoretical development would thus give a key tool for performing a coarse-grained description of protein dynamics in a comprehensive way.






\end{document}